# The Influence of Elastic Strain on Catalytic Activity Towards the Hydrogen Evolution Reaction

Kai Yan, Tuhina Adit Maark, Alireza Khorshidi, Vijay A. Sethuraman, Andrew Peterson,* and Pradeep R. Guduru*

School of Engineering, Brown University, Providence, RI 02912, USA

E-mail: *Andrew Peterson (andrew_peterson@brown.edu);
*Pradeep R. Guduru (pradeep_guduru@brown.edu)

**Abstract:** *Understanding the role of elastic strain in modifying catalytic reaction rates is crucial for catalyst design, but experimentally, this effect is often coupled with a ligand effect. To isolate the strain effect, we have investigated the influence of externally applied elastic strain on the catalytic activity of metal films towards the hydrogen evolution reaction (HER). We show that elastic strain tunes the catalytic activity in a controlled and predictable way. Both theory and experiment show strain controls reactivity in a controlled manner consistent with the qualitative predictions of the HER volcano plot and the d-band theory: Ni and Pt's activity were accelerated by compression, while Cu's activity was accelerated by tension. By isolating the elastic strain effect from the ligand effect, this study provides a greater insight into the role of elastic strain in controlling electrocatalytic activity.*

The full sequence of elementary steps in a general heterogeneous catalytic reaction involves adsorption of reacting species, dissociation and association of chemical species, transport on the surface, and desorption of the product species. [1a] The role of elastic strain in tuning catalytic reaction rates has evolved rapidly in the last decade and has initiated a significant re-evaluation of catalyst design.[1,2] Specifically, researchers have examined the role of misfit strain arising when a catalytically-active metal overlayer is epitaxially deposited on another metal substrate or when a shell metal is deposited around a core metal to form a core/shell nanoparticle.[1] In general, the misfit strain changes the width of the d-band through changes to the d-orbital interactions that are quite sensitive to interatomic spacing $r$, scaling as $r^{-5}$.[1b, 2] Changes in the d-band width modify reactivity of the strained surface by shifting the mean d-band energy (the "d-band center") relative to the Fermi energy, which influences the bonding and anti-bonding states of adsorbates and reactants on the metal surface.[3] The relationship between the adsorption energy and the d-band center was confirmed, for example, by calculations for CO adsorption on various surfaces,[1a] which in combination with the adsorbate scaling relations and the Brønsted-Evans-Polyani (BEP) relationship relates the d-band center to reaction rates and the Volcano Plot.[4]

Early experimental evidence of strain effects on phenomena related to catalysis was reported by Gsell et al.[5] who demonstrated that O adsorption on Ru could be enhanced under tensile strain. The lattice strain effect on HER activity was also studied by Wolfschmidt using a Pt-modified Au(111) catalyst; it was shown that monatomically high Pt islands on Au(111) have enhanced catalytic activity.[6] Other experimental work has involved core-shell nanostructures, which introduce mismatch strains in the outer shell, and also de-alloying to control misfit or surface strains.[7] Voiry et al.[8] reported chemically exfoliated $WS_2$ as efficient catalysts for $H_2$ evolution with low overpotentials and the enhanced electrocatalytic activity of $WS_2$ is due to the high concentration of the strained metallic octahedral phase in the as-exfoliated nanosheets. Strasser et al. studied PtCu@Cu core@shell system formed by de-alloying Pt-Cu nanoparticles,[9] and attributed increases in catalytic activity for oxygen reduction reaction (ORR) to the elastic strain in the de-alloyed shell. In their study, the strain in the shell was estimated indirectly through anomalous X-ray diffraction of nano-particles before and after de-alloying.[10] The saturation of activity with the estimated elastic strain and the absence of the expected Volcano plot were attributed to possible relaxation mechanisms in the shell. However, there are a few studies that separate the strain-effect from the ligand effect. Recently, Smetanin et al.[11] reported changes in the potential and current for predominantly capacitive processes in 20-nm-thick Au films supported on polyimide substrates subjected to cyclic loading under galvanostatic and potentiostatic conditions, which were explained in terms of the influence of strain on the surface capacitive processes. Importantly, Deng et al.[12] studied the influence of elastic strain in thin films of Au and Pt on polyimide on their electrocatalytic activity for HER. In these studies, a small oscillatory tensile load was applied to the catalyst and the response was quantified in terms of an electrocapillary coupling parameter. The authors assumed kinetic rate equations, based on a Heyrovsky limiting step and Langmuir-isotherm coverage behavior, to individually assign the strain dependence of the hydrogen adsorption enthalpy and the activation enthalpy.

The dynamic coupling between strain and reactivity provided a link between mechanics and adsorption to give a better understanding of reactivity.[12,13] More recently, Yang et al.[14] demonstrated a similar effect for ORR on a Pd-based metallic glass catalyst film under both tensile and compressive strains, which have opposite effect on catalytic activity. Du et al.[15] further confirmed the elastic strain on NiTi shape memory alloy in ORR, whereas the compressive strain enhanced the ORR activity and the strain can be influenced by the applied temperature. Sethuraman et al.[16] subjected thin Pt films on single crystal Si substrates to external straining while the films were participating in ORR through cyclic voltammetry (CV). They showed that a tensile strain of 0.7% resulted in ~15 mV reduction in the overpotential. In a computational study, Adit Maark et al.[17] studied HER over a Pd(111) surface under in-plane biaxial strain and examined the effect of hydrogen coverage.

Following up on the above investigations, here we report a systematic study of the effect of externally applied elastic strain on catalytic activity of pure metal films towards HER. Production of $H_2$ from water splitting is one of the most well-studied electrocatalytic reactions.[18] HER, which is the half-cell reaction responsible for the evolution of $H_2$, it is ideally suited for studying the effect of elastic strain as it is well established that the binding energy of H — expressed as the free-energy change of hydrogen adsorption ($\Delta G_H$)—is the primary predictor of the effectiveness of an electrocatalyst for HER, with an ideal catalyst having $\Delta G_H$ of about 0 eV.[19] Catalytic activity of various metals for HER is often represented by a volcano plot, which shows activity versus $\Delta G_H$. On such a plot, the peak of the volcano (maximum activity) occurs at $\Delta G_H$ of 0 eV. The activity decreases to the left of volcano peak due to stronger H binding ($\Delta G_H < 0$), the activity decreases to the right of the peak as well where H binding is weaker ($\Delta G_H > 0$).[20] According to the d-band model, compressive strains generally weaken the binding strength of adsorbates whereas tensile strains enhance it.[14, 21] Hence, we should expect compressive strain to weaken the

H binding energy on metals to the left of the volcano peak, moving them towards the peak and thus enhancing their activity towards HER; tensile strain should decrease activity. However, strain is expected to have the opposite qualitative effect on metals to the right of the volcano peak, *i.e.*, compressive strains should move those metals further from the volcano peak and decrease the HER activity; while tensile strains should increase activity. In this investigation, we choose three metals that span the volcano plot: Ni (left of the peak), Pt (near the peak) and Cu (right of the peak), which provide a representative set to investigate the effect of elastic strain and the predictions of the d-band model. The metal films are subjected to externally-applied compressive and tensile loading while they are participating in HER. Simultaneously, we performed density functional theory (DFT) calculations on Pt, Ni, and Cu(111) surfaces under various loading conditions to more precisely quantify the theoretical predictions, and further interpret the experimental measurements in terms of changes in the H binding energy.

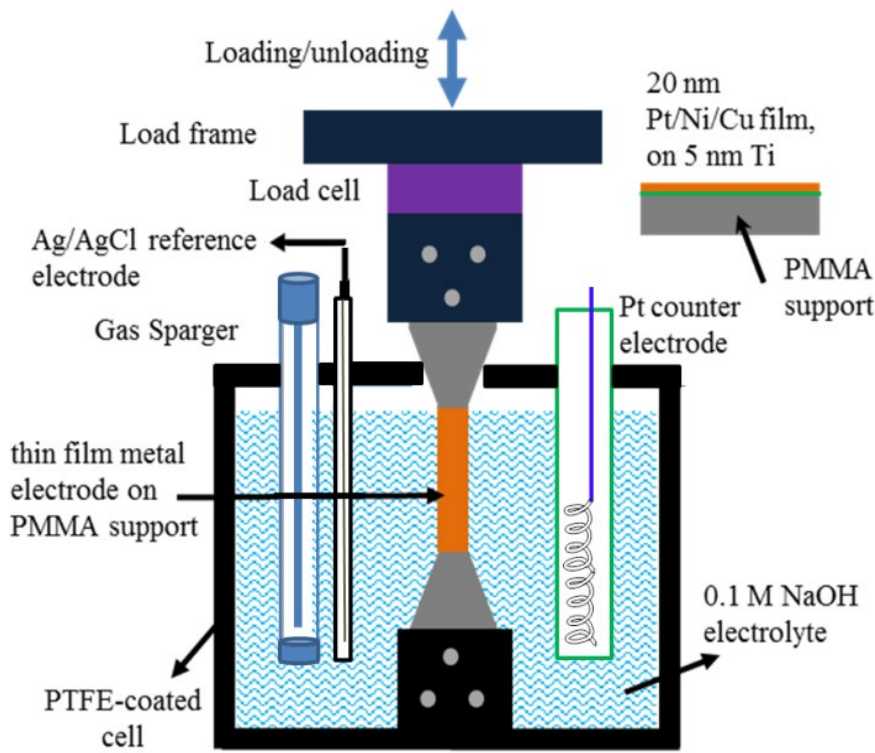

*Figure 1. Experimental setup to subject thin metal films to elastic straining by applying compressive or tensile loading on the PMMA substrates in a universal mechanical tensing machine.*

Thin films of the three metals under consideration (Pt, Ni and Cu) were synthesized by sputtering and/or e-beam evaporation on PMMA (polymethylmethacrylate) substrates. Details of the deposition process are provided in Supporting Information (SI). PMMA surfaces can be very smooth, which was confirmed by atomic force microscopy (AFM) for our samples, showing a mean absolute surface roughness, $R_a$, of ~3.8 nm. AFM measurements showed a roughness of 6 nm in the as-deposited metal films. As calculated by the quad triangle method from the AFM images, the actual surface area of the catalyst films has approximately 2% greater surface area than the projected area, and thus we use the geometric current density, relative to the unstrained catalyst, as an indicator of catalyst performance. Surface roughness mapping also revealed the as-deposited films to be continuous and homogeneous. Elastic strains were applied on the films by subjecting the PMMA substrates to uniaxial tensile and compressive loading. The thin films are expected to inherit the in-plane strain behavior of the PMMA substrate, as described in the SI. An electrochemical cell was designed to be integrated with a universal mechanical testing machine as shown in Figure 1. (The assembly and contact of the electrode are shown in Figure S3.) The elastic constants of PMMA substrates were independently measured (Young's modulus: 3.2 GPa, Poisson's ratio: 0.38, and yield strain: ~1.6 %) so that the applied load can be used to calculate the elastic strain. The Poisson's ratio of the PMMA was also assumed to dictate the response of the combined system in the electronic structure calculations described later. As shown in Figure 2, the activity of the films for HER was studied at each load value through cyclic voltammetry (CV) in 0.1 M NaOH electrolyte by monitoring the changes in the potential at specified values of HER current.

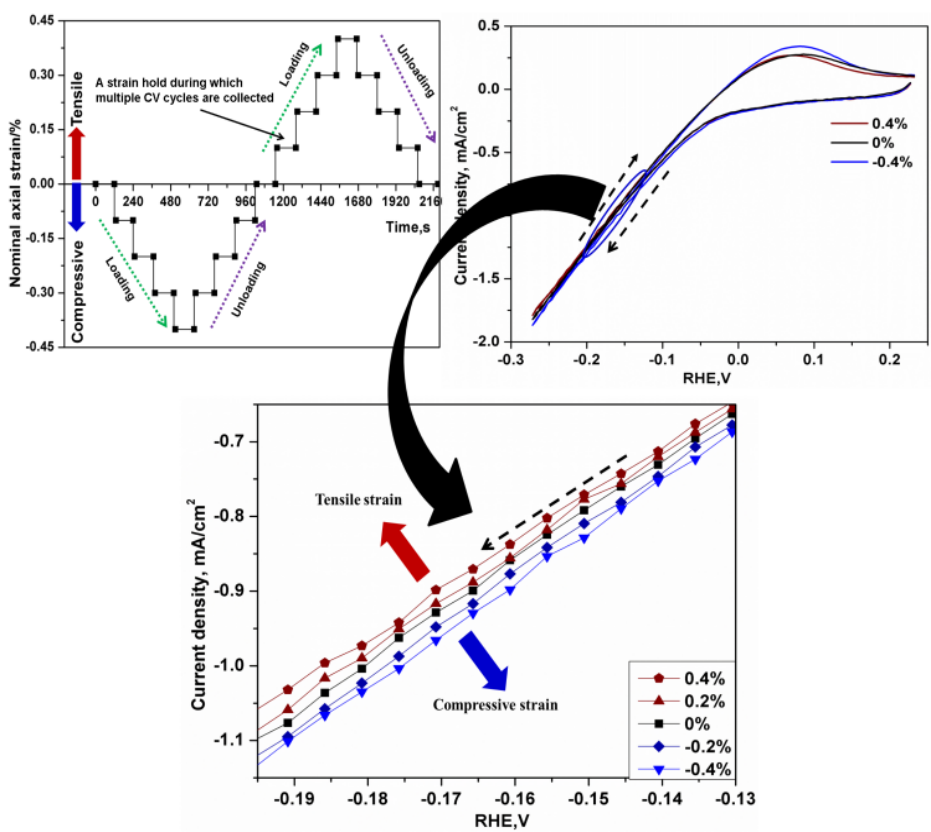

*Figure 2*. (a) Schematic illustration of the loading history on the PMMA substrates, showing progressively increasing compressive and tensile strains. The Pt films deposited on PMMA inherit the substrate strains. (b) Representative CVs obtained at strains of -0.4%, 0% and 0.4%, where negative values are compressive and positive values are tensile. (c) Magnified view of (b) in the HER region showing only the reduction sweep at 5 different strain values. Note the systematic shift in the CV curves with strain.

To study the effect of elastic strain on HER, the following procedures were adopted to the three studied thin-metal-film catalysts. Following assembly, the electrochemical cell was subjected to

30 CV scans at a sweep rate of 50 mV/s to ascertain that a steady state was established. The PMMA substrates were then subjected to a strain cycle as illustrated in Figure 2a. First, the sample was subjected to progressively higher compressive loads corresponding to nominal elastic strains (that is, percent elongation or compression) of -0.1%, -0.2%, -0.3% and -0.4%; the sample was held at each constant strain while five CV scans were collected.

During unloading, the sample was again held at the same nominal strains while CV scans were collected at each hold. Following the compressive cycle, the samples were subjected to progressively increasing tensile loads corresponding to nominal tensile strains of the same magnitude, followed by decreasing tensile loads (Figure 2a). The samples were held at nominal tensile strains of the same magnitude as above while CV scans were collected at each hold. Following the strain cycle, 30 CV scans were collected again to confirm that films remained stable and the response remained identical to that before the strain cycle. In addition, following the experiment, the films were examined by scanning electron microscopy (SEM) to ascertain that they did not undergo any mechanical damage such as cracking or delamination. AFM measurement of the film after electrochemical tests still showed a low roughness (Ra ~2.7 nm), suggesting the geometric current density continued to be an appropriate measure. Note that the strain changed the surface area by changing the bond lengths between atoms; for this reason, we did not expect the strain to affect the number of catalytic sites and reported all current densities normalized by the unstrained area. As calculated in the SI, changing the assumed area of the electrode under these small strains has no significant effect on the results. At each strain, the average of the five collected CVs was obtained and the scatter among them was used to determine the error estimate through standard error estimation procedures. Figure 2b shows CVs for Pt films collected at three strain values, 0%, 0.4% and -0.4%. With the introduction of strain (*e.g.*, at -0.4% nominal strain), a shift in the reductive overpotential of ~4 mV was observed.

To consistently assess the three metals, we chose a narrow region of the CV where hysteresis (capacitive) effects between the oxidative and reductive sweeps were absent; details on the choice of this region are in the SI. Figure 2c shows a detailed view of the CVs for five strain values (0%, ±0.2% and ±0.4%) during the reduction potential sweep. As seen in Figure 2c, a systematic shift of the CV curve as a function of strain is observed; compressive strain shifts the CVs to the right (higher current and lower overpotential), thus increasing the catalytic activity. Tensile strain had the opposite effect, *i.e.*, shifted the curves to the left, thus lowering the activity. Although the simplest interpretation of the potential shift in the CVs is in terms of changes in reaction kinetics at the surface, it is possible that the experiments are in a regime coupling reaction and diffusion. However, even in the coupled regime, the rate of diffusive transport becomes larger when the rate of reaction increases (and vice versa), that is, the shift of potential due to strain has the same direction. Thus, even in the coupled regime, the potential shifts in the CVs would continue to reflect changes in reaction kinetics; *i.e.*, the catalytic activity. However, it should be noted that this could make the comparison of potential shifts to DFT predictions (see below) more qualitative. The magnitude of the strain effect was slightly different for the three metals (Pt case in Figure 2, Ni and Cu in Supporting Information, Figure S5), but it was generally in the range of ~10 mV per 1% strain.

Experimental measurements of the CV shift for all three metals, *i.e.*, Pt, Ni and Cu are shown in Figures 3a-c. We have taken the convention of Deng *et al.*[12] of describing the strain in terms of areal deformation, $e = \Delta A / A_0$, which includes the elongation/compression in the loading direction as well as the response in the co-planar transverse direction, as dictated by the Poisson's ratio of the PMMA substrate. The shift of the CV to the right (*i.e.*, reduction in overpotential or increase in activity) was taken as positive and that to the left (increase in overpotential or lower activity) is taken as negative. Details are described in Supporting Information. Note that Figures 3a-c cycle as well as the scatter among multiple specimens.

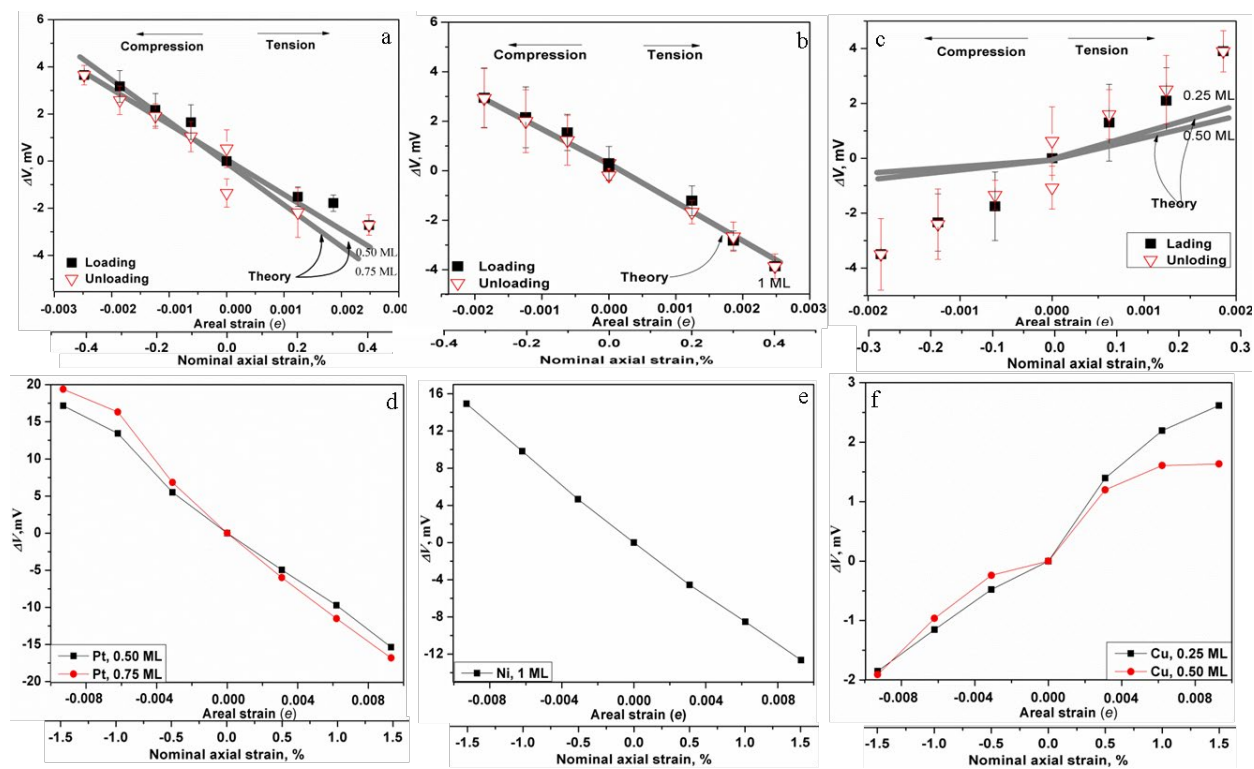

*Figure 3. CV-curve-shifts relative to the zero strain level as a function of applied elastic strain for (a) Pt, (b) Ni and (c) Cu. A positive potential shift denotes a reduction in overpotential and vice versa. Note the qualitative difference between Cu and the other two metals; compressive strain increases catalytic activity for Pt and Ni whereas the effect is opposite for Cu, which is on the other side of the volcano peak. The solid lines (gray color) on the experimental data represent the computational results from the bottom row of figures. The bottom row shows the potential shifts obtained computationally for (d) Pt(111), (e) Ni(111), and (f) Cu(111) surfaces at different hydrogen coverages as a function of in-plane predominantly uniaxial strain. Areal strain, $e = \Delta A / A_0$, where $\Delta A$ and $A_0$ are the changed area under nominal axial strain and the initially physical area of the electrode, respectively. The plots shown are average responses for straining along perpendicular in-plane directions. Note that the computationally accessible range of strain values is much larger than what was experimentally accessible.*

Figure 3 reveals a number of important features: (i) The voltage change present results from multiple experiments for each material; the error bars include the error estimates within each individual was approximately linear with applied elastic strain for all three metals in the experimentally accessed strain range. The magnitude of the slope of the relation was in the range of 10 mV per 1% strain for all three metals; For Pt, a nominal tensile strain of 0.4% ($e = 0.0025$) induced an increase of overpotential of 2.6 mV. The general trend is in agreement with previous findings in Pt films by Deng et al. [12], albeit with a higher magnitude. For the Ni case, a similar trend was also observed, while the inverted trend was observed in Cu film, whereas 0.4% tensile strain induced a reduction of overpotential with 3.9 mV; (ii) Compressive strain increased the HER activity of Pt and Ni and decreased that of Cu. In Pt case, the -0.4% compressive nominal strain led to a reduction in overpotential with 4.3 mV, while it caused an increase of the overpotential with -3.5 mV in Cu film. This was a striking observation and is in good agreement with the theoretical predictions also shown in the Figure 4, discussed further below. Our findings on the coupling trends in the Pt and Cu cases agree well with the work of Deng et al.; [12] however, the current findings for Ni may disagree with the predictions of Deng et al. [12] which may be illuminated in further investigations. (iii) These results clearly separate the strain effect from the ligand effect, thus establishing the magnitude of the former, which can provide a useful guideline for the design of nanostructured catalysts.

As noted earlier, the hydrogen binding energy is regarded as a good first descriptor of the HER activity of a transition metal surface. To provide a comparison between experiment and theory, we carried out density functional theory calculations of the hydrogen binding energy on fcc(111) surfaces for the same metals under the application of in-plane uniaxial loading in the range of ±1.5%, in a geometric manner designed to mimic the experimental conditions. We assumed the thin metal film experiences the same in-plane response, in terms of uniaxial strain and Poisson response, as that of the substrate.

For the loading configuration shown, the substrate experiences a strain of $\varepsilon$ in the axial direction (i.e. loading direction) and a transverse strain of $-\upsilon_{PMMA}\varepsilon$, where $-\upsilon_{PMMA}$ of 0.38 is the Poisson's ratio of the PMMA. In the direction normal to the plane, the catalyst was allowed to relax without constraint, as would be expected in the physical sample. Full details of the calculations are provided in the Supporting Information. At the atomic scale, each metal surface was constructed using a 2 × 2 non-orthogonal unit cell with four layers and 20 Å vacuum, employing the corresponding DFT bulk lattice constants, calculated as 4.021 Å for Pt, 3.552 Å for Ni, and 3.710 Å for Cu. Among the four layers normal to the surface, only the bottom one is fixed in the model, such that the metal can respond in the $z$ direction. (We employ the convention, Figure 4a, that the Cartesian $z$ direction is normal to the (111) surface.) The free energies for H adsorption ($\Delta G_H$) at 298 K are calculated via the computational hydrogen electrode model as:

$$\Delta G_H(\theta_H) = \Delta E_H(\theta_H) + \Delta E_{ZPE} - T\Delta S_H + \Delta G_V \qquad (1)$$

where $\Delta E_H(\theta_H)$ is the differential hydrogen adsorption energy:

$$\Delta E_H(\theta_H) = E(nH^*) - E((n-1)H^*) - \tfrac{1}{2} E(H_2) \qquad (2)$$

In the above equations $\theta_H$ is the H coverage (0.25-1 monolayer, ML). To compare with previous work, we also calculated "electrocapillary coupling coefficients", as defined in reference [13a], for H on Pt to be 1.7 V and 2.0 V for 0.5 ML and 0.75 ML, respectively; these are similar to the value reported in reference [13a] of 1.9 V for 1 ML. More details are provided in the Supporting Information. The free energy diagram for HER over unstrained M is depicted in Figure 4b at 0 V vs RHE. It can be seen that adsorption of H species (H*) on the surface is downhill in energy for the cases of Pt and Ni and the subsequent H$_2$ evolution is uphill, suggesting the latter is the potential limiting step. In contrast, for Cu(111), H* formation is uphill but the removal of H$_2$ is downhill; hence, this suggests that HER activity is limited by the first step on Cu. As discussed earlier, it has been predicted from theoretical calculations [21,22] that the peak in HER activity corresponds to a free energy ($\Delta G_H$) of 0 eV. Thus, increasing HER activity of Ni, as well as Pt, should require a weakening of their H binding energy and that of Cu would require a strengthening of the H adsorption energy, although we note that unstrained Pt is very close to the peak. The calculated variation of the free energy ($\Delta G_H$) of H adsorption over Pt, Ni, and Cu(111) surfaces as a function of the applied strain are shown in Figures 4c-e; in these figures $\Delta G_H$ has been averaged over the x- and y-directions of the unit cell (Figure 4a). In accordance with the d-band theory, compressive strain weakens H binding while tensile strain strengthens it over Pt, Ni, and Cu(111) surfaces at all coverages. The effect of H coverage is evident from the $\Delta G_H$ variation for Pt and Cu: as H coverages ($\theta_H$) increase H adsorption strength weakens at all strains considered here. It can be further deduced from Figures 4c-e that (i) $\Delta G_H$ of Pt(111) could reach values close to 0 eV at relatively high compressive strain of about -1.4%, (ii) $\Delta G_H$ of Ni(111) is strongly negative even at a high coverage of 1.0 ML and large compressive strains, and (iii) $\Delta G_H$ of Cu(111) remains much greater than 0 eV over the full strain range considered here.

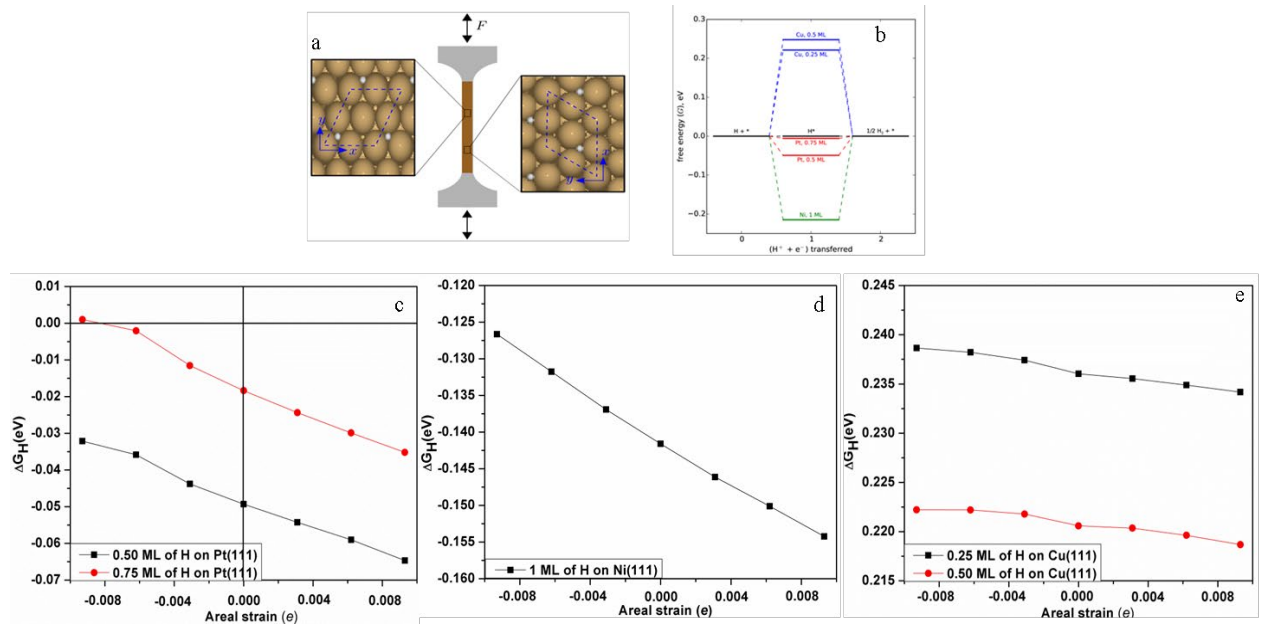

*Figure 4.* (a) A two-scale schematic of the M/PMMA sample (M = Pt, Ni, or Cu) under uniaxial loading. The M film is modelled as a polycrystalline material, where half of the M(111) surface has x-direction of the unit cell parallel to the loading axis, and the other half has x-direction of the unit cell perpendicular to the loading axis. The illustrated M(111) surfaces are 0.25 ML (monolayer) H-adsorbed at fcc sites. (b) Free energy diagram for HER over unstrained Pt, Ni, and Cu(111) surfaces at different surface coverage $\theta_H$. Variation of free energy of hydrogen adsorption ($\Delta G_H$) with in-plane uniaxial strain at different $\theta_H$ for (c) Pt(111), (d) Ni(111), and (e) Cu(111).

To make a quantitative connection to the experiments, we define the limiting potential $V_L$ as the electrode potential at which the free energy change of all unit steps in the reaction is less than or equal to zero, *i.e.*, the reaction becomes downhill in the free energy landscape.(See, for example, reference [23].) It can be seen from Figure 4b that $V_L$ is given by $-(\Delta G_H$ of the most uphill step)/$e$ where $e$ is the charge of an electron. Although the energy barriers for the unit steps are ignored, the difference between the equilibrium potential and the limiting potential has often been found to give an indication of the overpotential requirement for the catalytic reaction. Hence, the change in the limiting potential with strain, *i.e.*, $\Delta V = V_L$ (strained) $- V_L$ (unstrained) can be compared with the experimentally observed shift in the CV curves with strain. Figures 3(d-f) plot the computed $\Delta V$ with strain for the three metals (for different coverages) in the strain range of -2.5% to 1.5%. We note that the plotted results represent the average response corresponding to multiple loading directions in the (111) plane. The computational results within the experimental loading range were superposed on the experimental data in Figures 3a-c as gray solid lines. Despite the simplicity of the computational model, the agreement with the experimental data is remarkably good in all three cases. In particular the computations also capture the reversal in the strain effect on HER activity of Cu compared to that of Pt and Ni. This also suggests that the peak of the volcano plot is correctly placed to the right of Pt; that is, Pt binds H slightly too strongly.

In summary, we experimentally demonstrate that externally applied elastic strain can influence the HER activity of Pt, Ni and Cu thin-film catalysts in a predictable way, consistent with the predictions of the d-band model and with the hydrogen-evolution volcano. By separating the strain effect from the ligand effect, we demonstrate that the effect of strain on metals on opposite sides of the volcano peak is reversed, *i.e.*, compressive strain increases the catalytic activity of metals to the left of the volcano peak and reduces that of the metals to the right. Tensile strain has the opposite effect. The experimental observations match theoretical calculations qualitatively and quantitatively. The combined experimental and computational study presented here shows the power of separating the strain and ligand effects and offers new insights into the design of catalysts not only for HER, but also other electrocatalytic reactions of interest.

**Acknowledgements**

This work was supported by the U.S. Army Research Laboratory and the U.S. Army Research Office under the Multi University Research Initiative MURI (W911NF-11-1-0353) at Brown University (Program Manager: Dr. David Stepp). Electronic structure calculations were carried out at the Brown University Center for Computation and Visualization (CCV).

**References:**

[1] a) M. Mavrikakis, B. Hammer, J. K. Nørskov, *Phys. Rev. Lett.* **1998**, *81*, 2819-2822; b) J. R. Kitchin, J. K. Nørskov, M. A. Barteau, J. G. Chen, *Phys. Rev. Lett.* **2004**, *93*, 156801; c) W. Tang, G. Henkelman, *J. Chem. Phys.* **2009**, *130*, 234103.
[2] a) B. Hammer, Y. Morikawa, J. K. Nørskov, *Phys. Rev. Lett.* **1996**, *76*, 2141-2144; b) A. Ruban, B. Hammer, P. Stoltze, H. L. Skriver, J. K. Nørskov, *J. Mol. Catal. A* **1997**, *115*, 421-429; c) B. Hammer, J. K. Nørskov, *Surf. Sci.* **1995**, *343*, 211-220; d) B. Hammer, L. B. Hansen, J. K. Nørskov, *Phys. Rev. B* **1999**, *59*, 7413.
[3] L. A. Kibler, A. M. El-Aziz, R. Hoyer, D. M. Kolb, *Angew. Chem. Int. Ed.* **2005**, *44*, 2080-2084; *Angew. Chem.* **2005**, *117*, 2116–2120.

[4] a) F. Abild-Pedersen, J. Greeley, F. Studt, J. Rossmeisl, T. R. Munter, P. G. Moses, E. Skúlason, T. Bligaard, J. K. Nørskov, *Phys. Rev. Lett.* **2007**, *99*, 016105; b) T. Bligaard, J. K. Nørskov, S. Dahl, J. Matthiesen, C. H. Christensen, J. Sehested, *J. Catal.* **2004**, *224*, 206-217; c) A. A. Gokhale, J. A. Dumesic, M. Mavrikakis, *J. Am. Chem. Soc.* **2008**, *130*, 1402-1414.
[5] M. Gsell, P. Jakob, D. Menzel, *Science* **1998**, *280*, 717-720.
[6] W. Holger, B. Rainer, S. Ulrich, *J. Phys.: Condens. Matter* **2008**, *20*, 374127.
[7] J. Wu, P. Li, Y. T. Pan, S. Warren, X. Yin, H. Yang, *Chem. Soc. Rev.* **2012**, *41*, 8066-8074.
[8] D. Voiry, H. Yamaguchi, J. Li, R. Silva, D. C. Alves, T. Fujita, M. Chen, T. Asefa, V. B. Shenoy, G. Eda, *Nature Mater.* **2013**, *12*, 850-855.
[9] P. Strasser, S. Koh, T. Anniyev, J. Greeley, K. More, C. Yu, Z. Liu, S. Kaya, D. Nordlund, H. Ogasawara, M. F. Toney, A. Nilsson, *Nature Chem.* **2010**, *2*, 454-460.
[10] a) C. Cui, L. Gan, H. Li, S. Yu, M. Heggen, P. Strasser, *Nano Lett.* **2012**, *12*, 5885-5889; b) A. Bergmann, I. Zaharieva, H. Dau, P. Strasser, *Energy Environ. Sci.* **2013**, *6*, 2745-2755.
[11] M. Smetanin, Q. Deng, J. Weissmuller, *Phys. Chem. Chem. Phys.* **2011**, *13*, 17313-17322.
[12] Q. Deng, M. Smetanin, J. Weissmüller, *J. Catal.* **2014**, *309*, 351-361.
[13] a) J. Weissmüller in *Electrocatalysis: Theoretical Foundations and Model Experiments* (Eds.: R. C. Alkire, L. Kibler, D. M. Kolb, J. Lipkowski), Wiley-VCH, Weinheim, **2013**, pp.163-219; b) Q. Deng, D. H. Gosslar, M. Smetanin, J. Weissmüller, *Phys. Chem. Chem. Phys.,* **2015**, *17*, 11725-11731; c) L. Lührs, C. Soyarslan, J. Markmann, S. Bargmann, J. Weissmüller, *Scripta Mater.* **2016**, *110*, 65-69; d) Q. Deng, V. Gopal, J. Weissmüller, *Angew. Chem. Int. Ed.* **2015**, *54*, 12981-12985; *Angew. Chem.* **2015**, *127*, 13173-13177.
[14] Y. Y. Yang, T. Adit Maark, A. Peterson, S. Kumar, *Phys. Chem. Chem. Phys.* **2015**, *17*, 1746-1754.
[15] M. Du, L. Cui, Y. Cao, A. J. Bard, *J. Am. Chem. Soc.* **2015**, *137*, 7397-7403.
[16] V. Sethuraman, D Vairavapandian, MC Lafouresse, TA Maark, N Karan, S. Sun, U. Bertocci, A. Peterson, G. R Stafford, P. Guduru *J. Phys. Chem. C* **2015**, *119*, 19042-19052.
[17] T. A. Maark, A. A. Peterson, *J. Phys. Chem. C* **2014**, *118*, 4275-4281.
[18] a) O. Khaselev, J. A. Turner, *Science* **1998**, *280*, 425-427; b) M. Symes, L. Cronin, *Nature Chem.* **2013**, *5*, 403-409; c) R. Michalsky, Y. Zhang, A. Peterson, *ACS Catal.* **2014**, *4*, 1274-1278; d) Y. Zheng, Y. Jiao, M. Jaroniec, S. Z. Qiao, *Angew. Chem. Int. Ed.* **2015**, *54*, 52-65; *Angew. Chem.* **2015**, *127*, 52-66.
[19] a) R. Parsons, *Trans. Faraday Soc.* **1958**, *54*, 1053-1063; b) J. K. Nørskov, T. Bligaard, A. Logadottir, J. Kitchin, J. Chen, S. Pandelov, U. Stimming, *J. Electrochem. Soc.* **2005**, *152*, J23-J26.
[20] a) S. Trasatti, *J. Electroanal. Chem. Interf. Electrochem.* **1972**, *39*, 163-184; b) J. K. Nørskov, J. Rossmeisl, A. Logadottir, L. Lindqvist, J. R. Kitchin, T. Bligaard, H. Jónsson, *J. Phys. Chem. B* **2004**, *108*, 17886−17892.
[21] a) J. Greeley, T. F. Jaramillo, J. Bonde, I. Chorkendorff, J. K. Nørskov, *Nature Mater.* **2006**, *5*, 909-913; b) Q. Deng, J. Weissmüller, *Langmuir* **2014**, *30*, 10522-10530; c) N. Huber, R. N. Viswanath, N. Mameka, J. Markmann, J. Weißmüller, *Acta Mater.* **2014**, *67*, 252-265; d) R. N. Viswanath, J. Weissmüller, *Acta Mater.* **2013**, *61*, 6301-6309.
[22] a) J. K. Norskov, C. H. Christensen, *Science* **2006**, *312*, 1322-1323; b) J. Greeley, M. Mavrikakis, *Surf. Sci.* **2013**, *540*, 215-229.
[23] A. Peterson, F. Abild-Pedersen, F. Studt, J. Rossmeisl, J. K. Nørskov, *Energy Environ. Sci.* **2010**, *3*, 1311-1315.

# Supporting Information

**The influence of elastic strain on catalytic activity towards the hydrogen evolution reaction**

Kai Yan, Tuhina Adit Maark, Alireza Khorshidi, Vijay Sethuraman,
Andrew Peterson,* Pradeep R. Guduru*

School of Engineering, Brown University, Providence, RI 02912, USA

*Corresponding authors: pradeep_guduru@brown.edu; andrew_peterson@brown.edu

This document contains further details on both the experimental and computational methodologies.

**Experimental details**

*Materials.* Platinum (Pt) sputtering target (1.00" dia. x 0.063" thick, 99.99% purity), Copper (Cu) evaporation pellets (1/4" dia. x 1/2" length, 99.999% purity) and Nickel (Ni) evaporation pellets (1/4" dia. x 1/2" length, 99.995% purity) were purchased from Kurt J. Lesker. Acetone (ACS reagent, 99.5%) was purchased from Sigma-Aldrich and sodium hydroxide (NaOH, 96.7%) was purchased from Macron Chemicals.

*Thin-film deposition.* Thin metal films of Pt, Cu and Ni were deposited by a Modular Thin Film Deposition System (Lesker 18) by sputtering and electron beam deposition on PMMA substrates (base vacuum 5 x $10^{-6}$ Torr), at deposition rates of 0.9 Å/s to 1.2 Å/s. The thickness of all films in this study was 20 nm, which was controlled by a thickness monitor.

*Characterization of the deposited thin films.* Surface mapping was measured by Zygo optical surface profiler capable of measuring feature heights ranging from < 2 nm up to 20000 μm, it was found that all the deposited thin films display homogeneous surfaces. X-ray diffraction (Bruker D8 Discover using monochromatic Cu Kα radiation at 40 KV and 40 mA) was carried out on the deposited thin films (Figure S1). No measurable changes in the diffraction patterns were seen after strain cycling.

***Electrochemical measurements***. Electrochemical studies were carried out in a standard three electrode system controlled by a Metro (Autolab PGSTAT 30) electrochemistry workstation. In all experiments, an Ag/AgCl reference electrode was employed and Pt coiled wire was used as the counter electrode. 0.1 M NaOH solution was used as the electrolyte, which was prepared by dissolving the calculated amount of NaOH with the conductivity of 0.056 s/m water (at 25 °C) to form 0.1 M solution. The electrolyte was sparged with ultra-high purity argon (99.999%).; all potentials reported in this paper are converted to the pH-independent reversible hydrogen electrode (RHE) scale. All experiments were repeated three times to confirm reproducibility. The electrochemical measurements were conducted in a three-electrode cell which was fabricated to be integrated with an Instron 4505 universal mechanical testing machine; the experimental setup is illustrated in Figure 1 of the main text. The design and the connection of the as-obtained metal film as working electrode are shown in Figure S3. This setup allows *in-situ* measurements of electrochemical responses while the specimen is under uniaxial tension and compression. The scan rate of all electrochemical measurements was 50 mV/sec.

The cell was constructed such that the electrolyte-gas interface defined the top of the active electrochemical area; in this way, the geometric differences between active catalyst area in strained and unstrained cases were minimized. To simplify the analysis, we normalized all geometric current densities by the unstrained value (which further implicitly assumes that the number of active sites did not change with strain). Here, we examine the sensitivity of our results to this assumption, using data from case of Pt in unstrained versus at 0.4% nominal strain and comparing three different geometry assumptions. Our unstrained electrode height ($L_1$) was approximately 25 mm and its width ($W_1$) was approximately 6.5 mm. Therefore, its unstrained area was $A_1 = L_1W_1 = 162.5$ mm$^2$, which by our working assumption (Assumption #1) was used to normalize the current for all strain cases. At our maximum reported loading of +0.4%, we can first (Assumption #2) calculate that the height expands to $L_2 = 25.1$ mm (0.4% elongation), and the width will shrink by the Poisson ratio of the PMMA substrate (0.38) to $W_2 = 6.5 \times (1 - 0.004 \times 0.38) = 6.49$ mm$^2$. Under the assumption that both these dimensional changes lead to a change in catalytic area, this leads to $A_2 = L_2W_2 = 162.9$ mm$^2$. Under another scenario (Assumption #3), which matches the physical description most closely, we assume that the experimental setup results in the length being unchanged as the top edge of the electrocatalyst is defined by the electrolyte-gas interface. Therefore, in this scenario $L_3 = L_1 = 25$ mm and $W_3 = W_2$; therefore, $A_3 = L_3W_3 = 162.25$ mm$^2$. Choosing a typical point on the CV of -0.166 V, we recorded a current density at 0% loading of -0.899 mA/cm$^2$. For the case of +0.4% nominal strain, under Assumptions #1, #2, and #3, respectively, the strained current density would be reported as -0.871, -0.869, and -0.872 mA/cm$^2$. That is, all three assumptions lead to -0.87 mA/cm$^2$ when reported to the two significant digits to which we have confidence, as opposed to -0.90 mA/cm$^2$ for the unstrained case. To a good approximation, we are insensitive to the method of normalizing area (for the case of these rather small strains), and thus we choose the simplest case of Assumption #1 in this study.

*Voltage change measurements*. In order to measure the potential shift due to strain, the difference between the reduction portion of the CVs corresponding to the strained and unstrained cases was obtained in a suitable current density range. The current density range was chosen such that the response was purely Faradaic in nature (corresponding to $H_2$ evolution electrocatalysis) as opposed to capacitive current. This was identified by the region of the CV in which the reductive (negative-going) and oxidative (positive-going) sweeps overlapped, rather than exhibiting hysteresis characteristic of a capacitive process. We chose this region first for the Pt catalyst, then for consistency chose the same distance from the visible "knee" on the CV of approximately -0.07 V. This strategy was employed to minimized capacitive effects; to attempt to minimize convolutions from mass transport, we chose the potential window we studied to be small (approximately 40 mV) to avoid extending the analysis into regions of large current density. However, in practice the curves exhibited the same shift also at the more negative potentials. Further information on the geometry of the cell is described above. For the case of Pt, $\Delta V$ was measured in the current density range of -0.7 to -0.3 mA/cm$^2$ at increments of 0.025 mA/cm$^2$ and the average of these values is taken as the potential shift for the corresponding applied strain. The corresponding current density ranges for Ni and Cu are -2.25 to -1.0 mA/cm$^2$ and -0.175 to -0.0375 mA/cm$^2$, respectively. The error bars shown in Figure 3 include the spread in the $\Delta V$ variation within the current range described above, as well as the sample-to-sample variation.

**Computational methodology**

The hydrogen evolution reaction was examined over Pt, Ni, and Cu(111) surfaces under application of an in-plane uniaxial strain. Each surface was constructed using a 2 × 2 unit cell with four layers and 20 Å vacuum employing the corresponding DFT bulk lattice constants. Introducing more metal layers is assumed to not change adsorption energies significantly, as has been previously investigated by Mavrikakis et al.[1] and Skúlason et al.[2]., and trends (differences) in binding energies can be expected to converge faster than the binding energies themselves.[3] Strain applied in one direction leads to a relaxation in the other two orthogonal directions. By assuming that the thin metal film experiences the same in-plane mechanical deformation as that of substrate, the amount of transverse relaxation was calculated from Poisson's ratio of the PMMA substrate, and was explicitly included while constructing the surface unit cell. The top three layers in the unit cell (and the adsorbates) were allowed to relax, forcing the bottom layer to the spacing of the bulk crystal (at appropriate strain). The exposed facet was not assumed to have a bias to its alignment with respect to the strain; that is, it is assumed to have uniformly random distribution of lattice directions. This assumption is reflected in the model by taking the average of the two cases in which the uniaxial strain was applied along the *x*- and *y*- directions of the unit cell; convention for *x*- and *y*- axes of the unit cell is adopted according to Figure 4a (shown by blue color) of the manuscript, with the *z*-direction being normal to [111] surface.

All density functional theory (DFT) calculations were performed using the planewave DACAPO electronic structure code with ionic manipulations in the Atomic Simulation Environment (ASE) [4], converging Kohn-Sham iterations to within a $10^{-5}$ eV variation in energy as well as $10^{-4}$ and $10^{-3}$ variations in density and occupation of states, respectively. To improve electronic convergence, the occupation of Kohn-Sham states was smeared according to a Fermi-Dirac distribution with a thermal energy $k_B T$=0.1 eV, with standard extrapolation of electronic energies to $k_B T = 0$ eV before the application of statistical-mechanical temperature effects, described below. Electronic wave functions were expanded in a plane-wave basis set, and inner electrons were represented by Vanderbilt ultrasoft pseudopotentials.[5,6] In order to ensure the obtainment of converged results, fairly large values of 450 eV and 500 eV were used for the plane-wave cut-off and density cut-off, respectively. To approximate the exchange and correlation of electrons, the revised Perdew-Burke-Ernzerhof (RPBE) density functional of Hammer, Hansen, and Nørskov has been used, which was developed to be more accurate than PBE for the calculation of binding energies of adsorbates on transition-metal surfaces [7]. For the cases of Pt and Ni, the Brillouin zone integration was carried out using a 4 × 4 × 1 Monkhorst-Pack k-point grid for a system of 2 × 2 atoms in the unit cell [2, 8]. For the case of Cu, however, a finer *k*-point grid was necessary for the obtainment of converged energies, as has been also pointed out by Sakong and Groß [9]; here we used 10 × 10 × 2. In order to cancel out the net surface dipole density, standard dipole corrections were included at the farthest point in the vacuum in our calculations as implemented in the DACAPO calculator. In the case of Ni, spin polarization was taken into account; all other calculations were taken as spin-paired.

It is widely assumed that HER proceeds via the following mechanism in alkaline medium:

$$H_2O + e^- + * \rightarrow H^* + OH^- \quad (1)$$

$$2H^* \rightarrow H_2 + * \quad (2)$$

Alternatively the second step can be:

$$H_2O + e^- + H^* \rightarrow H_2 + * + OH^- \quad (3)$$

Similar to HER mechanism in acidic medium, the main entity in the above steps is the H adsorbed species, H*. Therefore, the free energies for H adsorption ($\Delta G_H$) can be used as a primary descriptor of HER activity. $\Delta G_H$ at 298 K is calculated via the computational hydrogen electrode (CHE) model [3] as:

$$\Delta G_H(\theta_H) = \Delta E_H(\theta_H) + \Delta E_{ZPE} - T\Delta S_H + \Delta G_V \quad (4)$$

where $\Delta E_H(\theta_H)$ is the differential hydrogen adsorption energy:

$$\Delta E_H(\theta_H) = E(nH^*) - E((n-1)H^*) - \tfrac{1}{2} E(H_2) \quad (5)$$

In Eq. (5), $n$ = 1-4 H atoms as coverage increases from 0.25 to 1.0 monolayer (ML). A pH correction to $\Delta G_H(\theta_H)$ is not invoked as the reference is the reversible hydrogen electrode (RHE). The correction for the electrical potential $\Delta G_V = -e\,V$, where $e$ is the (positive) charge of an electron (Faraday's constant divided by Avogadro's number) and *V* is the potential relative to 0 V RHE. We note that the CHE model considers only the energetics of the stable elementary thermodynamic states (adsorbed hydrogen), and not barriers between them. However, barriers are assumed to follow linear free-energy relations with respect to endstate energetics, and thus the simple voltage shifts implied by the CHE model can be expected to correlate with experimental measurements. The zero-point energy of an atomic configuration can be obtained from a normal-mode vibration analysis according to $E_{ZPE} = h/2 \ \Sigma_i^{DOF} \nu_i$, where the summation runs over the vibrational degrees of freedom in the system. When a hydrogen atom is adsorbed to the surface, we assume that the frequencies of vibration of the metal atoms of the slab are not affected by the adsorbed hydrogen. Therefore, when calculating the difference between zero-point energies of the system before and after hydrogen adsorption, the terms corresponding to frequencies of vibration of slab atoms cancel out, and we have

$$\Delta E_{ZPE} = 0.5 h \left( \sum_{i=1}^{3n} \nu_i - \sum_{j=1}^{3(n-1)} \nu_j - \nu_0 \right) \quad (5)$$

where $\nu_i$ is the frequency of vibration of *n* adsorbates, $\nu_j$ is the frequency of vibrations of *n* - 1 adsorbates, and $\nu_0$ is the frequency of stretch of hydrogen molecule (calculated here as 3291.3 cm$^{-1}$). After frequencies of vibration are found, entropy of a canonical ensemble can be calculated from

$$\frac{S(T)}{N} = k_B \sum_{i=1}^{3n} \left( \frac{h\nu_i}{k_B T (e^{h\nu_i/k_B T}-1)} - \ln\left(1 - e^{-h\nu_i/k_B T}\right) \right) \quad (6)$$

When calculating differential entropy change $\Delta S_H$, the entropy of gas-phase hydrogen atom was assumed as $S_H = \frac{1}{2} S_{H_2} = 6.77 * 10^{-4}$ eV/K, the entropy of gas-phase hydrogen at room temperature. We have further assumed that entropy and zero-point energy contributions stay the same for different values of strain, and therefore they have been calculated for zero-strain cases only.

Hydrogen adsorption was considered at the most favored adsorption sites. For H coverages ($\theta_H$) of 0.25, 0.75, and 1.0 ML (monolayer, where 1 ML corresponds to 1 H atom per surface metal atom), these are *fcc* sites [8, 10], which we confirmed in the current study. For an H coverage of 0.5 ML, we found the lowest energy configuration on Pt to be with all hydrogen atoms at *fcc* sites, but for Ni and Cu surfaces with half of the hydrogen atoms at *fcc* sites and the other half at *hcp* sites (as reported in the literature for Ni[11]), which was taken into account in the calculations. To estimate the reactive hydrogen coverage ($\theta_H$), we started with the experimental work of Markovic *et al.* [10] where adsorption isotherms of underpotential deposition of H on Pt(111) in 0.05 M $H_2SO_4$ relating $\theta_H$ to potential are reported. According to their study, at a temperature of 303 K, $\theta_H$ of 0.25 and 0.68 ML occurs at RHE potential ($U$) of about 0.23 and 0 V, respectively. In a subsequent DFT study, Skúlason *et al.*[2] predicted that a low $\theta_H$ of 0.25 ML on Pt(111) corresponds to a moderate positive potential of $U = +0.18$ V. When a full monolayer of hydrogen was formed, $U$ was close to zero. Therefore, in our DFT study, we considered $\theta_H$ of 0.5 and 0.75 ML for Pt(111). Moreover, our calculated binding energies of a single hydrogen atom on unstrained M(111) show that hydrogen adsorption is much stronger on Ni and considerably weaker on Cu as compared to Pt. Therefore, in the proceeding analysis, we assumed that at $U = 0$ V, hydrogen coverage over Ni and Cu would be, respectively, higher and lower than Pt, leading to assuming $\theta_H = 1$ ML for Ni (as has been also reported by the reference [10]) and $\theta_H = 0.25$ and 0.5 for Cu.

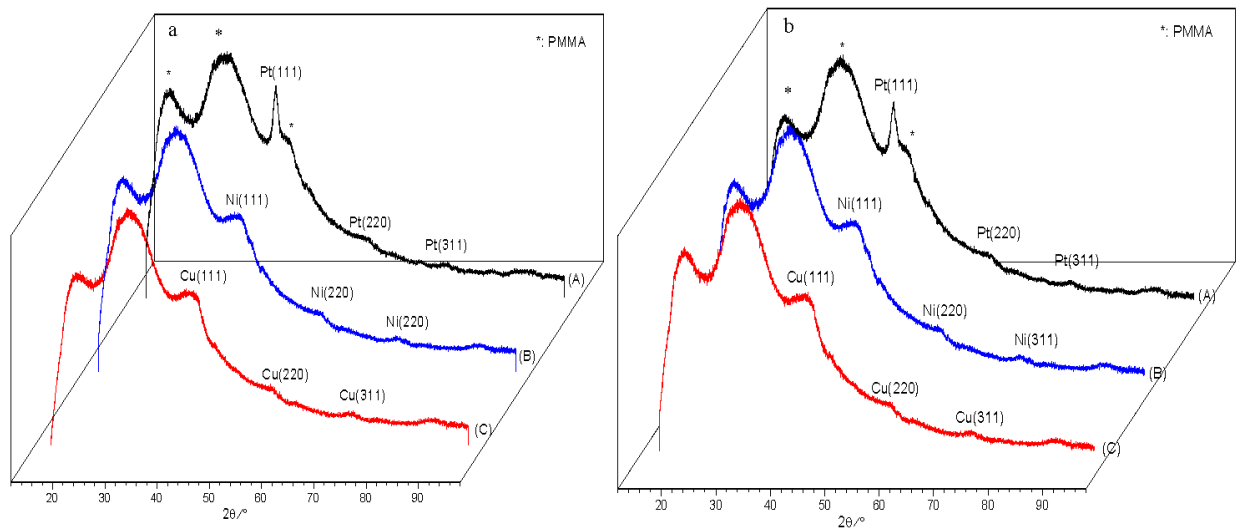

**Figure S1.** XRD patterns of the fresh (a) and spent (b) metal thin films. The XRD patterns before and after the experiment were nearly identical, with no measurable changes in the film texture.

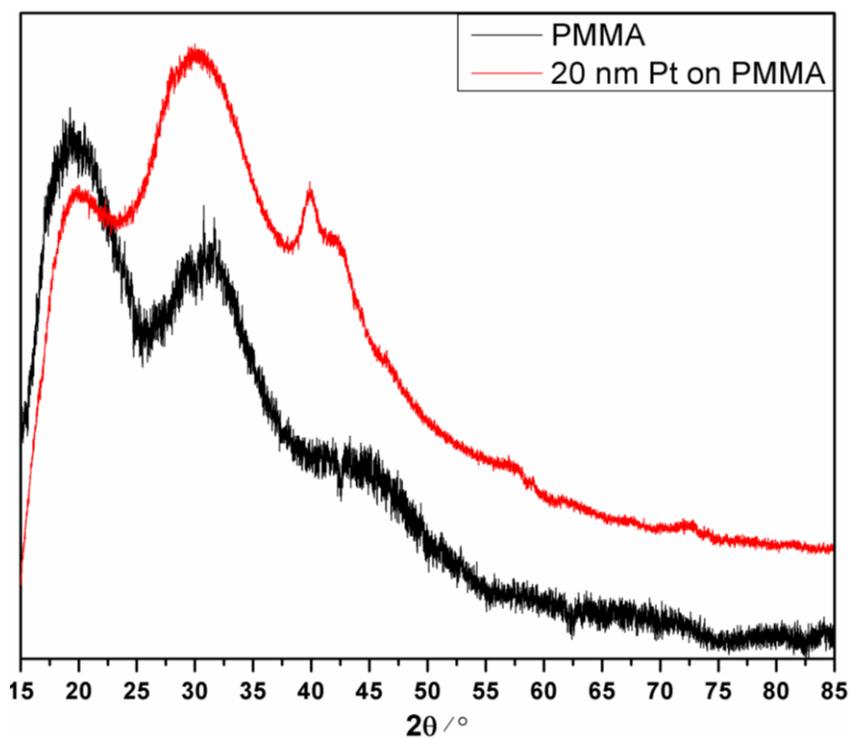

**Figure S2**. The diffraction pattern of PMMA substrate and the as-obtained 20 nm Pt films on PMMA substrate.

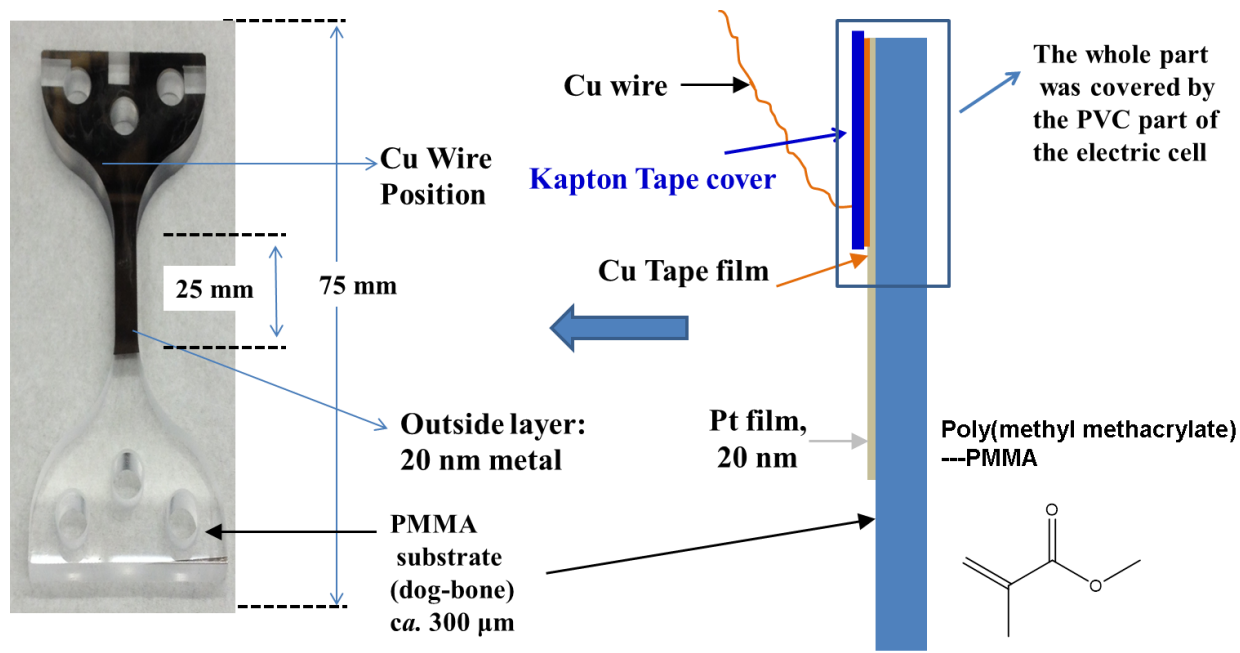

**Figure S3**. The assemble and contact of metal films as the working electrode.

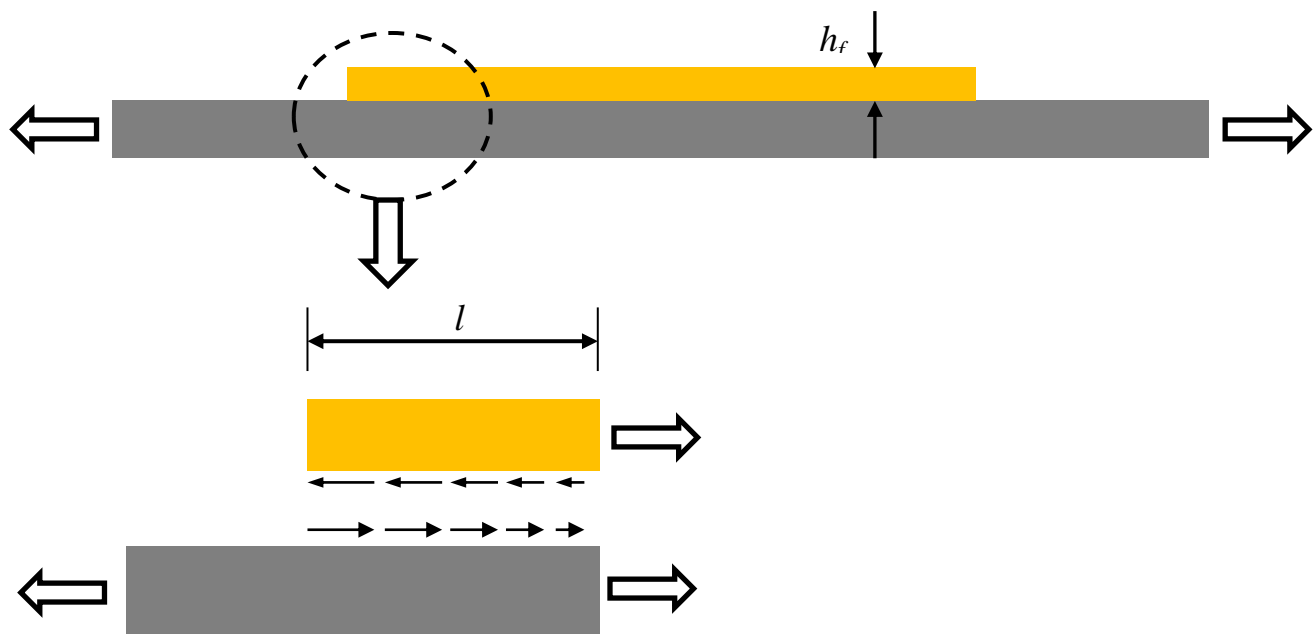

**Figure S4**. (top) Schematic illustration of a film bonded on a substrate, which is subjected to in-plane loading and elastic strain as shown. (bottom) Magnified view of the film edge, in which the arrows represent the shear stress that is induced along the interface. The length of the arrows represents the shear stress magnitude; note that the magnitude of the shear stress approaches zero over a distance $l$ from the film edge. Note that in-plane stress in the film at the left edge is zero

(it is a free-surface). However, as the as we move away from the edge, normal stress develops in the film, which is necessary to maintain equilibrium by counteracting the interfacial shear stress. At a distance $l$ from the edge, the shear stress approaches zero and the in-plane stress in the film reaches a steady state value, which corresponds to the elastic strain in the substrate.[12]

**Strain transfer**

Subjecting films to tensile or compressive stress by straining the substrates on which they are deposited is a common and widely used technique.[13-19] For an elastic film on an elastic substrate, when the substrate is strained, the film inherits the substrate strain beyond a transition length $l$ from its edge, as shown above. So, except in these transition regions, the film strain is equal to that of the substrate.[12] The transition length $l$ can be estimated from previous works.[20]

$$l \sim 10\, h_f/k$$

where

$$k = \frac{E_s}{1-v_s}\frac{1-v_f}{E_f}$$

$E_s$ and $E_f$ are the Young's moduli of the substrate and the film respectively; $v_s$ and $v_f$ are the Poisson's ratios of the substrate and the film respectively. For the combination of PMMA substrate and metal films, $k \sim 1/100$, hence $l \sim 1000\, h_f$. In our experiments, $h_f \sim 20$ nm, which results in a transition length of $l \sim 20$ μm, which is negligible compared to the in-plane dimensions of the film of about 6.5 mm x 25 mm. Since the area of the transition zone is a negligible fraction of the total film area, the entire film can be considered to be under the same elastic strain as that of the substrate.

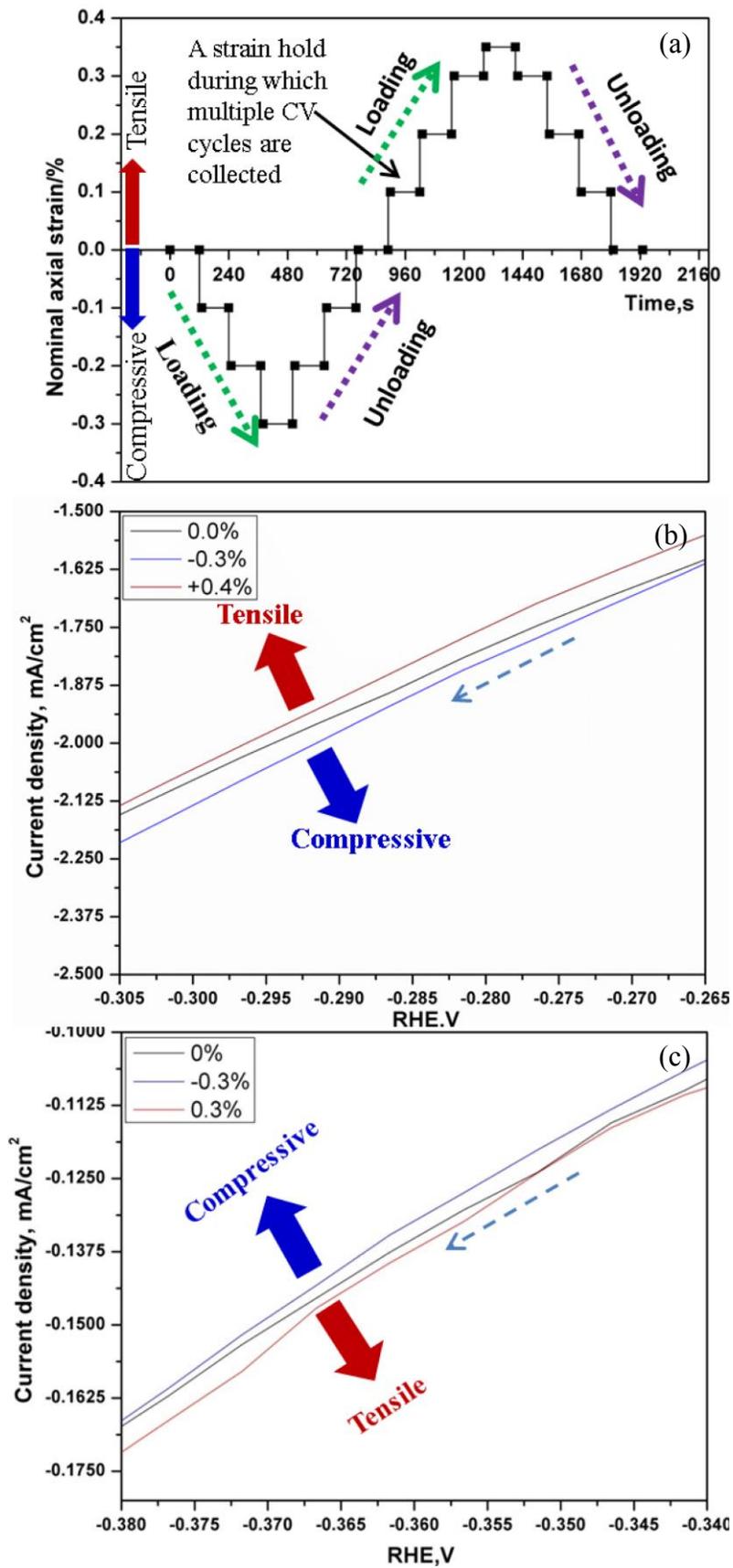

**Figure S5**. (a) Schematic illustration of the loading history on the PMMA substrates, showing progressively increasing compressive and tensile strains. (b) Magnified view of Ni thin film in the HER region showing only the reduction sweep at 3 different strain values. (c) Magnified view of Cu thin film in the HER region showing only the reduction sweep at 3 different strain values. Note the systematic shift in the CV curves with strain.

**References**

[1] M. Mavrikakis, B. Hammer, J. K. Nørskov, *Phys. Rev. Lett.* **1998**, *81*, 2819.

[2] E. Skúlason, G. S. Karlberg, J. Rossmeisl, T. Bligaard, J. Greeley, H. Jónsson, J. K. Nørskov, *Phys. Chem. Chem. Phys.* **2007**, *9*(25), 3241-3250.

[3] A. J. Medford, J. Wellendorff, A. Vojvodic, F. Studt, F. Abild-Pedersen, K. W. Jacobsen, T. Bligaard, J. K. Nørskov, *Science,* **2014**, *345*, 197-200.

[4] S. R. Bahn, K. W. Jacobsen, *Comput. Sci. Eng.* **2002**, *4*, 56-66.

[5] D. Vanderbilt, *Phys. Rev. B* **1990**, *41*, 7892-7895.

[6] J. K. Nørskov, J. Rossmeisl, A. Logadottir, L. Lindqvist, J. R. Kitchin, T. Bligaard, H. Jónsson, *J. Phys.Chem. B* **2004**, *108*,17886−17892.

[7] B. Hammer, L. B. Hansen, J. K. Nørskov. *Phys. Rev. B* **1999**, 59, 7413.

[8] T. A. Maark, A. Peterson, *J. Phys. Chem. C* **2014**, *118*, 4275-4281.

[9] S. Sakong, A. Groß, *Surf. Sci.* **2003,** *525*, 107-118.

[10] N. M. Marković, B. N. Grgur, P. N. Ross, *J. Phys. Chem. B* **1997**, *101*, 5405-5413.

[11] J. Greeley, M. Mavrikakis, *Surf. Sci.* **2003**, *540* (2)*,*215-229.

[12] L. B. Freund, S. Suresh, *Thin film materials: stress, defect formation and surface evolution*. Cambridge University Press, **2004**, P223-237.

[13] N. Lu, X. Wang, Z. Suo, J. Vlassak, *Appl. Phys. Lett.* **2007**, *91*(22), 221909.

[14] H. J. Jin, J. Weissmüller, *Science*, **2011**, *332*(6034), 1179-1182.

[15] M. Smetanin, Q. Deng, J. Weissmüller, *Phys. Chem. Chem. Phys.* **2011**, *13*(38), 17313-17322.

[16] Q. Deng, M. Smetanin, J. Weissmüller, *J. Catal.* **2014**, *309*, 351-361.

[17] M. Smetanin, D. Kramer, S. Mohanan, U. Herr, J. Weissmüller, *Phys. Chem. Chem. Phys.* **2009**, *11*(40), 9008-9012.

[18] J. Lohmiller, R. Baumbusch, O. Kraft, P. A. Gruber, *Phys. Rev. Lett.* **2013**, *110*(6), 066101.

[19] Y. Yang, T. A. Maark, A. Peterson, S. Kumar, *Phys. Chem. Chem. Phys.* **2015,** *17*, 1746-1754.

[20] L. B. Freund, S. Suresh, *Thin film materials: stress, defect formation and surface evolution*. Cambridge University Press, **2004**, P225.